\documentstyle[aaspp4]{article}
\begin{document}
\lefthead{Elmegreen}
\righthead{Phases and Structures of Interstellar Gas}

\title{Phases and Structures of Interstellar Gas}

\author{Bruce G. Elmegreen\altaffilmark{1}}
\altaffiltext{1}{IBM Research Division, T.J. Watson Research Center,
P.O. Box 218, Yorktown Heights, NY 10598, USA, bge@watson.ibm.com}

\begin{abstract}The thermal and chemical phases of the cool component of
interstellar gas are discussed. Variations with galactocentric radius
and from galaxy to galaxy are mostly the result of changes in the
ambient interstellar pressure and radiation field.  Interstellar
structure that is hierarchical or fractal in the cloudy parts and has
large and connected empty regions between these clouds is probably the
result of turbulence.  Such structure opens up the disk to the
transmission of OB star light into the halo, and it provides for a
diffuse ionized component that tapers away gradually from each dense
HII region.  Fractal cloud structure may also produce the cloud and
clump mass functions, and perhaps even the star cluster mass function.
\end{abstract}

to be published in "The Physics and Chemistry of the Interstellar Medium,"
3rd Cologne-Zermatt Symposium, ed. G. Winnewisser, J. St\"utzki, and
V. Ossenkopf, Aachen: Shaker-Verlag, 1999, in press

\section{Phases of Interstellar Gas}

Interstellar gas in the solar neighborhood has a variety of thermal
and chemical phases, including cold atomic gas inside or mixed with
cold molecular clouds
(e.g., Sato \& Fukui 1978), cool and warm atomic gas on the envelopes of
molecular clouds (Chromey et al. 1989; Andersson \& Wannier 1993;
Wannier et al. 1993) and in the diffuse, low density medium (Kulkarni
\& Heiles 1987), cold molecules in dark and dense clouds (Combes 1991)
and in translucent clouds (Magnani, Blitz \& Mundy 1985),
warm molecules in dense clouds near embedded luminous stars (e.g., 
Cesaroni et al. 1994),
dense photo-ionized gas in the neighborhoods of O and B-type stars,
low-density ionized gas between these stars (Walterbos \& Braun 1994; 
Reynolds 1995;
Ferguson et al. 1996), and hot, shock-heated gas near supernovae and
windy stars (Ostriker \& McKee 1988). 

There is also considerable variety in the types of molecules that are
present inside dark clouds, ranging from primarily H$_2$ in translucent
clouds (Spitzer \& Jenkins 1975; Magnani et al. 1998), with a possible
trace of PAH or long-chain molecules (Leger \& Puget 1984; 
Tulej et al. 1998),
to a mixture of complex molecules in dense clouds (e.g., Langer et al.
1997; see reviews by van Dishoeck and Thaddeus, this conference).

The thermal and chemical states of the cold and cool gas have such a
great variety because the clouds shield themselves to different levels from
photodissociative radiation and radiant heating. This makes the 
molecular abundance and
temperature depend on density, column density, metallicity, age, and
local radiation field, all of which vary from cloud to
cloud and with radius in the Galaxy.  The nature of cloud {\it structure}
does not change this much, because it is largely the result of turbulence,
self-gravity, and local explosions, which seem to act the same way everywhere. 

Here we discuss the dominant processes that affect the molecular and
thermal states of the gas, and we review the some of the aspects of
cloud structure that are likely to arise from turbulence. 

\subsection{Cloud Self-Shielding and Molecule Formation}

Interstellar clouds shield themselves from photo-dissociative
radiation when the molecular formation rate (in cm$^{-3} s^{-1}$) 
integrated over the cloud
radius exceeds the molecular dissociation flux (in cm$^{-2} s^{-1}$) 
from incident starlight
(Jura 1975; Federman, Glassgold, \& Kwan 1979; van Dishoeck \& Black
1988; Sternberg 1989; Hollenbach \& Tielens 1997). 
This means that clouds determine their own molecular abundances
for any particular radiation field, depending on their density, column density,
and metallicity. Dense clouds with
only moderate column densities can have the same H$_2$/H ratio as lower
density clouds with larger column densities. As a result, {\it regions with
larger pressures tend to be more molecular}, i.e., higher pressures lead
to higher densities and greater self-shielding at all cloud masses. 

One implication of this result is that most galaxies have a radial
gradient in the molecular fraction outside the nuclear or barred region
because the pressure is lower at larger radii (Elmegreen 1993; Honma,
Sofue, \& Arimoto 1995). The total (turbulent+magnetic+thermal)
interstellar medium pressure
varies approximately as $P\approx\left(\pi/2\right)G\sigma_g\sigma_T$
for gas column density $\sigma_g$ and total gas+star column density
in the gas layer, $\sigma_T$.   Typically, $\sigma_g\propto\sigma_T$, so
$P\propto\sigma_g^2$. This means that as $\sigma_g$ decreases exponentially
with radius in a typical spiral galaxy, the total pressure decreases with
radius too, and it has half the scale length. This corresponds to a rather
rapid decrease of interstellar pressure with galactocentric radius. 

Radial pressure gradients lead to the appearance of
``molecular rings'' in galaxies that have inner disk cutoffs from a bar
or bulge. Molecular rings are not true rings, like galactic resonance
rings, but only molecular emission concentrations that result from
smooth exponential disks with inner cutoffs. The rapid decrease of
molecular fraction with increasing radius beyond the ``ring'' results
primarily from large-scale pressure changes.

High latitude clouds, shells, spiral-arm dust lanes, and other pressurized
fronts should also have large molecular fractions compared to other
clouds with the same column density. A good example of the influence of pressure on
molecule formation is shown by the nearby L1457 cloud, which has H$_2$
primarily in the southern part, where it was recently
compressed (Moriarty-Schieven, Andersson \& Wannier 1997).

The {\it thermal} state in clouds is generally related to the {\it molecular} state,
at least in the solar neighborhood, because once a cloud begins to
shield itself in the H$_2$ lines, it also becomes optically thick from dust. Then it
removes its principle quiescent heat source, which is the photoejection
of electrons off grains (de Jong 1977; Draine 1978; Bakes \& Tielens
1994). Molecular clouds need not always be cold, though. Different
photons are involved with photodissocation, which is limited to spectral
lines, and photoejection, which is a continuum process. Besides, density
and column density can vary separately. Consequently, there are warm
molecular clouds in the diffuse phase, i.e., high density, low 
column-density 
clouds (Spitzer \& Cochran 1973).  These clouds tend to dominate
the diffuse cloud population in the inner (high pressure) regions of bright
galaxies (Polk et al. 1988; Honma, Sofue, \& Arimoto 1995). There are
also diffuse or dense molecular clouds in the high pressure regions of
{\it dim} galaxies that are so cold they do not emit much CO radiation
(Allen et al. 1995). 
Such temperature variations for molecular gas inversely affect the
conversion factor from integrated CO linewidth to molecular hydrogen.  High
temperature molecular clouds have proportionally less molecular matter
per unit CO emission than low temperature molecular clouds. 

\begin{figure} 
\vspace {2. in}
\includegraphics{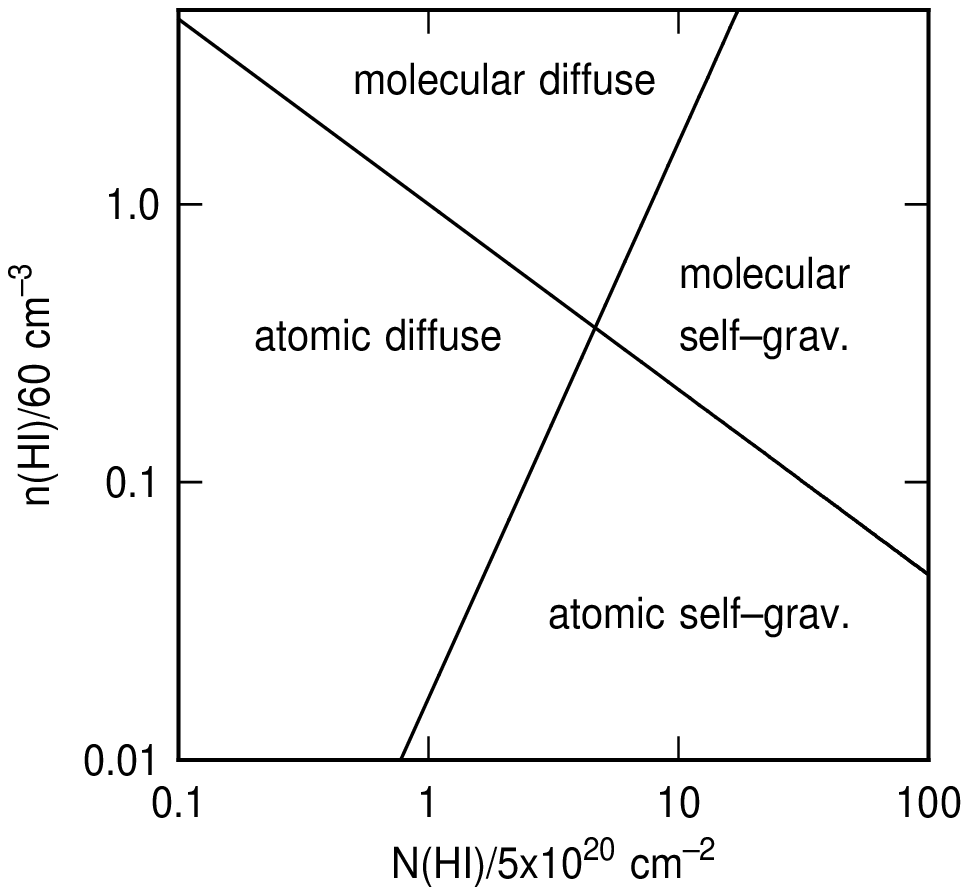}
\caption{Schematic diagram of interstellar cloud types on a density-column density
plane. Clouds turn molecular at high column density or at high density because
of self-shielding. Self-gravitating clouds require a sufficiently high
column density for a given ambient pressure, so they tend
to occur to the right in this diagram. Diffuse clouds require a low
column density, so they are on the left. The transition point depends on pressure
and radiation field and varies around the Galaxy. Figure from Elmegreen (1995).}
\label{fig:nN} \end{figure}

Figure \ref{fig:nN} shows a schematic diagram of the cold and cool
interstellar cloud phases in the parameter space that is defined by
density and column density. The rising line that separates diffuse from
self-gravitating gas comes from the relation \begin{equation}
{{P}\over{G\sigma^2}}=60 \left({{n}\over{60\;{\rm cm}^{-3}}}\right)
\left({{N}\over{5\times10^{20}\;{\rm cm}^{-2}}}\right)^{-2}
\left({{c}\over{2\;{\rm km}\;{\rm s}^{-1}}}\right)^{2}=1\;
\end{equation} for velocity dispersion $c$. This is the threshold for
strong self-gravity in a cloud. The falling line that separates
molecular from atomic gas comes from the equation \begin{equation}
\left({{Z/Z_0}\over{\phi/\phi_0}}\right) \left({{n}\over{60\;{\rm
cm}^{-3}}}\right) \left({{N}\over{5\times10^{20}\;{\rm
cm}^{-2}}}\right)^{2/3}=1 \label{eq:mole}
\end{equation} for metallicity $Z$ and
radiation field $\phi$, normalized to solar neighborhood values. This
relation is based on the balance between the H$_2$ formation rate (on
dust, hence the $Z$ dependence) along a column of gas, $\propto ZnN$,
and the photodissociation rate per unit area, which is proportional to
the absorbed photon flux, $\phi(N)\propto \phi N^{1/3}$. The column
density dependence in $\phi(N)$ accounts for extra absorption in the
line wings as $N$ increases (Federman, Glassgold, \& Kwan 1979).

Equation \ref{eq:mole} was considered in more detail by 
Sternberg (1989), who included a temperature dependence
for the molecule formation rate and dust extinction.  Sternberg
also approximated the 
column density dependence of $\phi(N)$ as $\propto N^{1/2}$ for pure
linewing absorption.  According to Sternberg, dust dominates 
molecules for the absorption of light in the shielding layer
when $\alpha G>>1$, which translates to 
\begin{equation}\alpha G\sim{{260\;{\rm cm}^{-3}\phi/\phi_0}\over
{n\left(\left[T/10\;{\rm K}\right]\left[Z/Z_0\right]\right)^{1/2}}}>>1.
\label{eq:sternberg}\end{equation} 
(This assumes Sternberg's molecule formation rate, $R$, and 
dust cross section, $\sigma$, both scale with metallicity, $Z$.)
If we scale the Federman, Glassgold \& Kwan (1979) result with
$\phi\propto N^{1/2}$ instead of $N^{1/3}$ and add a $T^{1/2}$ dependence
to the molecule formation rate (from the thermal speed at which H atoms
impact grains), then 
the equation of detailed balance implied by equation \ref{eq:mole}, 
$ZnNT^{1/2}\propto\phi(N)$, gives
a molecular column density for self-shielding of 
$N\propto \left(\phi/nZ\right)^2/T$,
and a corresponding dust opacity $\tau\propto ZN\propto
\left(\phi^2/n^2ZT\right)$.  The square root of this opacity is
essentially the quantity $\alpha G$ in Sternberg (1989), 
to within a factor of 2.
The column density at this
opacity limit is approximately $5\times10^{20}$ cm$^{-2}$ for standard
interstellar extinction, which means
that points to the right of $N(HI)/5\times10^{20}$ cm$^{-2}$
at high density $n$
in figure \ref{fig:nN} are primarily shielded by dust, and points to the
left are {\it self}-shielded by molecules. 
In both cases, the clouds are molecular, as indicated. 

Low density clouds are generally atomic except
at very high column density, at which point they may turn molecular
if the conditions are right. 
However, real clouds with very low densities are not likely to turn
molecular at high column density if their spatial extent is so
large that they include many field stars inside them. In that case,
there is no proper boundary for self-shielding. Such clouds will still
make a transition from diffuse to self-gravitating at moderate column
densities, however. This explains why the largest HI clouds in
spiral galaxies can be self-gravitating, and conversely, why
the largest self-gravitating clouds are often atomic 
(Elmegreen \& Elmegreen
1983; 1987; Rand 1995).

Figure \ref{fig:nN} also suggests that
high density clouds are generally molecular, except at very low column
density. When the density is high because of a high pressure, molecular
clouds can be diffuse. The internal density can also be high in a region
with a low ambient pressure provided the cloud is self-gravitating; this
makes the cloud molecular again.

In the Solar neighborhood, most clouds make a transition from atomic and
diffuse at low column density to molecular and self-gravitating at high
column density, with relatively little total cloud mass in the form of
molecular diffuse gas or atomic self-gravitating gas. However, in the
spiral arms of our Galaxy, atomic self-gravitating gas is prevalent in
the form of giant HI cloud complexes ($10^7$ M$_\odot$), which have
relatively small molecular cores ($10^5-10^6$ M$_\odot$) that are also
self-gravitating. Such clouds are often regularly spaced
along the spiral arms. In other galaxies such as M51, these $10^7$ M$_\odot$
spiral arm clouds are mostly molecular (Rand 1993). This is a sensible  result
for M51 because it has a relatively high total gas column density, and
therefore a high pressure ($P\sim G\sigma^2$). 

\begin{figure} 
\vspace {2. in}
\includegraphics{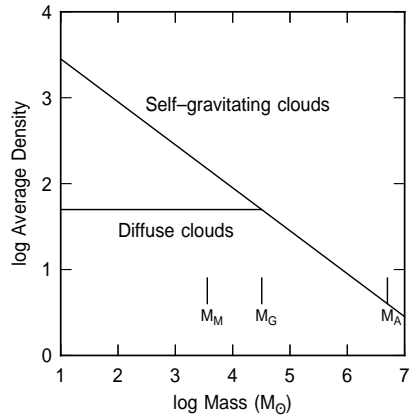}
\caption{Schematic diagram of interstellar cloud types on a density-mass parameter
plane. Self-gravitating clouds satisfy the density-mass relation from Larson
(1981) and occur for all masses, while the diffuse clouds, which are not
self-gravitating, have about constant density and occur only at small
mass.  The transition mass $M_M$ is where diffuse clouds turn from atomic
into molecular at the local pressure and radiation field; $M_G$ is where
diffuse clouds turn self-gravitating for local conditions, and $M_A$
is where molecular self-gravitating clouds turn atomic at higher mass
because of their low density. Figure from Elmegreen (1993).}
\label{fig:nM} \end{figure}

Figure \ref{fig:nM} makes these same points in a different way, showing
the densities of diffuse and self-gravitating gas as functions of cloud
mass. The Larson (1981) relation between density and mass is used for
the self-gravitating clouds. It comes from the two equations $P\sim
GM^2/R^4$ and $a^2\sim GM/\left(5R\right)$. 

At low mass, both diffuse and self-gravitating states are possible, as
in the usual two-fold solution to the virial equation with external
pressure: \begin{equation} 3Ma^2-{{3GM^2}\over{5R}}=4\pi R^3P.
\label{eq:virial} \end{equation} In this equation, for masses less than
$M=1.8a^4\left(G^3P\right)^{-1/2}$, 
there are stable states with low density and negligible
self-gravity, bounded by external pressure, and there are stable states
with high density and strong self-gravity, bounded by gravity, all at
the same mass. Similarly, in the interstellar medium, there are low mass
diffuse clouds as well as low mass self-gravitating clouds, such as Bok
globules. 

At high mass, equation \ref{eq:virial} has only one stable solution and
that is with self-gravity dominant. Similarly in the interstellar
medium, most massive clouds are strongly self-gravitating. This mass limit
changes when a magnetic field is present, but only slightly if the field
has the equipartition energy density (Mouschovias \& Spitzer 1976).

Also shown in figure \ref{fig:nM} are short lines that indicate transitions
from atomic to molecular phase as the mass increases ($M_M$), from
a mixture of non-self-gravitating and self-gravitating
clouds to pure self-gravitating clouds ($M_G$), and from
molecular back to atomic gas ($M_A)$. The origin of this latter
transition, which occurs at around $10^7$ M$_\odot$ for normal pressure,
is that the density of a massive cloud is so low as a result of the high
velocity dispersion and the requirement of pressure equilibrium, that the
cloud can no longer shield itself from photodissociative radiation. The
fact that the cloud is strongly self-gravitating does not matter.
Examples of such clouds are again the giant HI complexes in spiral arms.
At higher pressures, this transition from molecular to atomic
self-gravitating clouds occurs at higher masses. 

\subsection{Thermal States of Interstellar Gas}

Two thermal phases of atomic gas, cool and warm, can
co-exist in the Solar neighborhood at the {\it same} pressure and
radiation field because of inflections in the thermal cooling rate as a
function of temperature (Field, Goldsmith \& Habing 1969). 
At temperatures of around 100 K,
ionized Carbon and neutral Oxygen
cool the gas by collisions with neutral H, electrons, and, for OI,
ionized H (Wolfire et al. 1995). These collisions are very
strong coolants at temperatures near the energy levels of
CII and OI, which are 94 K and 228 K, easily balancing the photoelectric
heating rate that scales primarily with density. Thus the gas is thermally 
stable around 100 K. At lower temperatures, the CII and OI cooling transitions
are barely excited, but the cooling rate does not drop
much because the CI fine structure lines, which
have lower excitation temperatures, become important coolants.  Thus the gas remains
stable below 100 K too. At higher temperatures and lower densities with the
same pressure, the collision rates and cooling rates decrease, leading to
an unstable region until $T\ge5000$K, at which point rapid cooling from
Ly$\alpha$ emission and electron recombination onto grains makes
the gas stable again (Wolfire et al. 1995).

\begin{figure} 
\vspace {3.9in}
\includegraphics{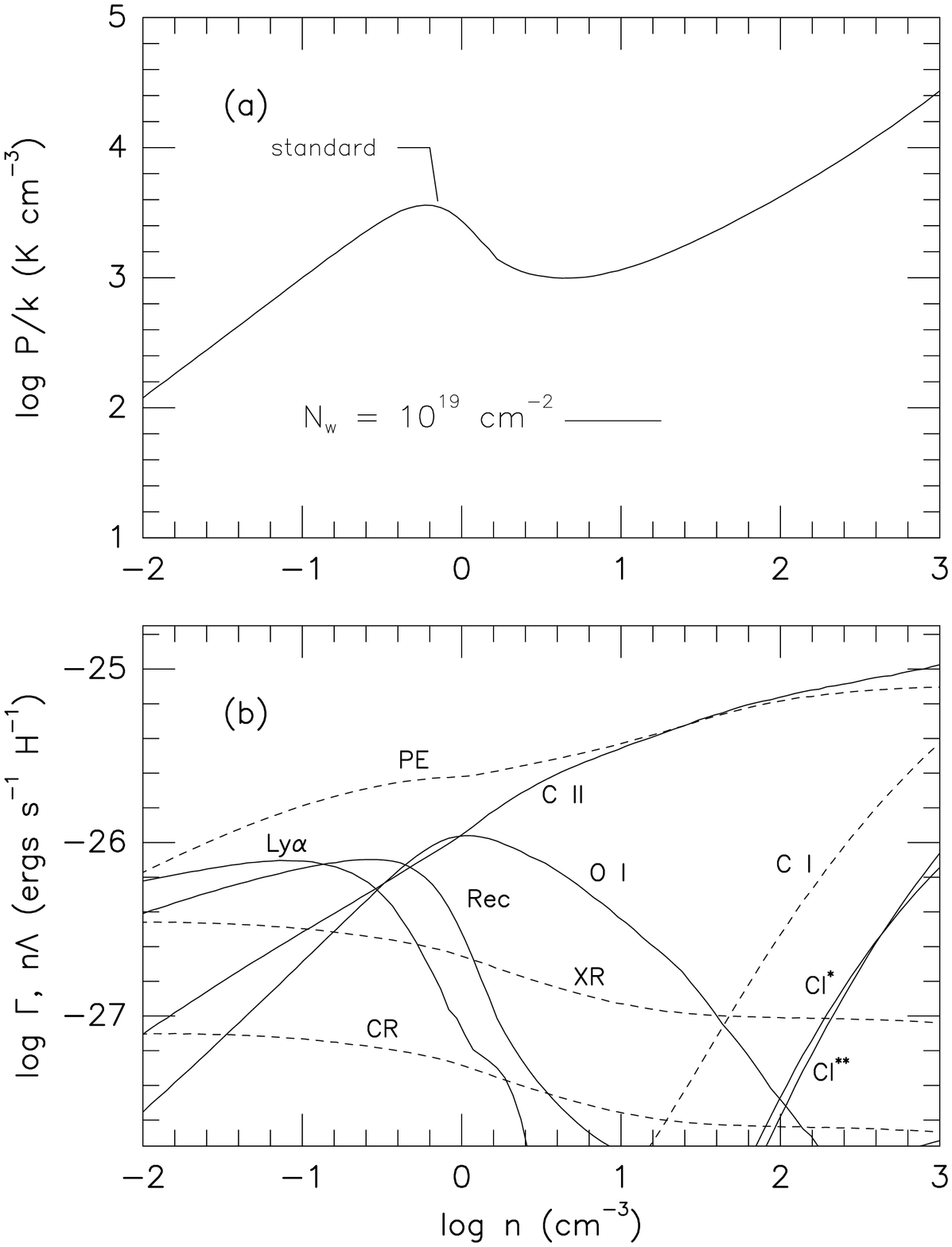}
\includegraphics{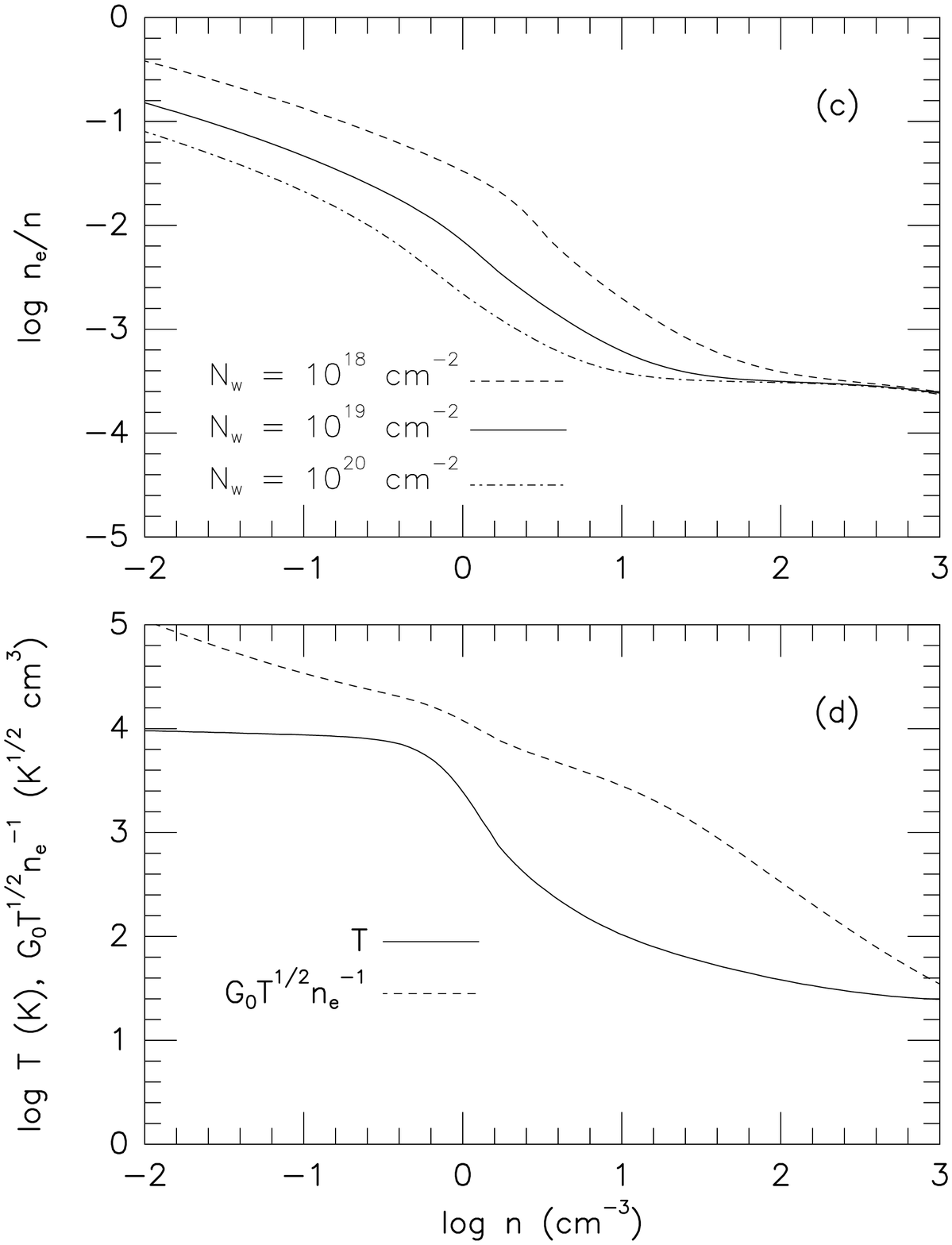}
\caption{(top left) The thermal pressure is shown as a function
of density for interstellar gas in thermal equilibrium. Only the rising
parts of this curve are in stable equilibrium, and these produce
the cool and warm neutral components locally. (lower left)
The heating (dashed) and cooling rates 
for neutral gas are shown as functions of density. In the upper and
lower right, the electron
fraction, temperature, and ionization parameter are shown. All
figures are from Wolfire et al. (1995).}
\label{fig:wolf} 
\end{figure}

Figure \ref{fig:wolf}
shows the thermal pressure, heating and cooling rates, ionization fraction, 
temperature, and ionization parameter ($G_0T^{1/2}/n_e$ for
radiation field $G_0$ scaled to the value in the Solar
neighborhood, temperature $T$, and electron density $n_e$) for the ``standard''
model in Wolfire et al. (1995).  

The local coexistence of two stable thermal states should not apply everywhere in a
disk galaxy. In the inner parts, where the pressure is very high and the
ambient radiation field is only moderately high, there may not be a warm
atomic phase at all, and a high fraction of the atomic gas can be
cool. Conversely, at large galactocentric radii, the pressure drops
precipitously but the radiation field only a small amount, so there
comes a point, at about 4 exponential scale lengths, where most of the
HI becomes warm (Elmegreen \& Parravano 1994; Braun 1997). 

Local variations in the atomic state can arise because of variations in
the radiation field and pressure. These variations often follow a spiral
density wave. For example, most of the gas between the spiral arms is at
low pressure, so it should be highly atomic, mostly diffuse, and warm. A
population of low-mass, self-gravitating molecular clouds typically remains as
well, from the previous arm. When this gas shocks to form a dust lane in
the arm, the density and opacity to starlight go up in the atomic gas,
the temperature drops, and the gas should turn molecular, even
ultra-cold molecular. But if massive star formation occurs inside or
near this shocked gas, the enhanced radiation can warm it up again and
dissociate it, forming a concentration of cool or warm HI near the
resulting OB association, as observed in M83 
(Allen, Atherton \& Tilanus 1985; Tilanus \& Allen 1993),
M51 (Tilanus \& Allen 1989), 
M100 (Rand 1995), and M81 (Allen et al. 1997). Cold dense molecular
material can coexist with this warm atomic gas if the molecules are in
strongly self-gravitating clouds; the gravity preserves their high
density and self-shielding even in the presence of strong radiation
fields. 

The pressure, radiation field, and metallicity vary both systematically
and randomly from place to place and time to time in any one galaxy, and
from galaxy to galaxy along the Hubble sequence. What applies to the
Solar neighborhood will not generally apply elsewhere, and what typifies
the Milky Way can be relatively rare in galaxies with different Hubble
types. Nevertheless, most of the variations that have been observed in
other galaxies can
be explained as a consequence of the simple rules given above. 

The hotter phases of interstellar gas require local excess energy
sources such as hot stars and supernovae. These sources are relatively
rare in a typical galactic disk (not in a nuclear starburst disk,
though), but each commands such a wide domain of influence that the
locally heated regions can sometimes overlap with each other, making the hot
gas pervasive (Cox \& Smith 1974; Salpeter 1976; McKee \&
Ostriker 1977). The domain of influence increases at
lower density, so a single massive star at a large distance off the
galactic plane can ionize a large volume in the halo. For
example, an O5 star in a gas with a halo-type density of 0.03 cm$^{-3}$
can ionize out to 3 kpc distance. 

\section{Structure of Interstellar Gas}

\subsection{Hierarchical Structure in Clouds}

The origin of structure in interstellar gas is far less tangible than
the origin of thermal and chemical states, because some of the
structure is caused by supersonic turbulence, and no one
understands yet how this works. Other structures, such as shells
(Tenorio-Tagle \& Bodenheimer 1988) and spiral shocks (Roberts 1969),
are somewhat easier to understand because they are regular and well
defined. Turbulent structures, on the other hand, are irregular,
boundary-free, and extremely intricate.

For example, a single molecular cloud, such as the Orion cloud, probably
contains
a total number of dense tiny clumps and cloud pieces (each with masses
of $M<10^{-2}$ M$_\odot$ in the $10^5$ M$_\odot$ Orion cloud) that
exceeds the total number of all other Orion-type clouds in all of the
galaxies out to and including the Virgo cluster (considering
$\sim10^3-10^4$ Orion-type clouds per spiral galaxy the size of the
Milky Way). Obviously there are far too many pieces of structure in a
typical molecular cloud to catalog or map at the present time, and far
too many to simulate on a computer. Yet these tiny pieces are very
important: they probably contain most of the mass in the cloud, as
demonstrated by the generally high excitation densities for many
molecular transitions (e.g., Falgarone, Phillips, \& Walker 1991; Lada,
Evans \& Falgarone 1997), and therefore control most of the chemistry,
and ultimately, star formation. How should we proceed to understand
interstellar matter in the face of such complexity (Scalo 1990)?

One of the most fundamental properties of turbulent clouds seems to be
the arrangement of gas into hierarchical structures that are fractal or
multi-fractal, with increasing density and local volume filling factor on
smaller scales. Scale-free structures like this make a certain amount of
sense for turbulent gas because the velocities that continuously form
them have a scale-free nature as well, reminiscent of the Kolmogorov
velocity law in incompressible laboratory turbulence. Scale-free
velocities should make scale-free structures, as long as thermal
pressure and gravity are not important structuring agents as well. When
gravity is important, the gas becomes centrally condensed and builds up
a pressure gradient. Most star-forming clouds are centrally condensed
already.

\begin{figure} 
\vspace {4.in}
\includegraphics{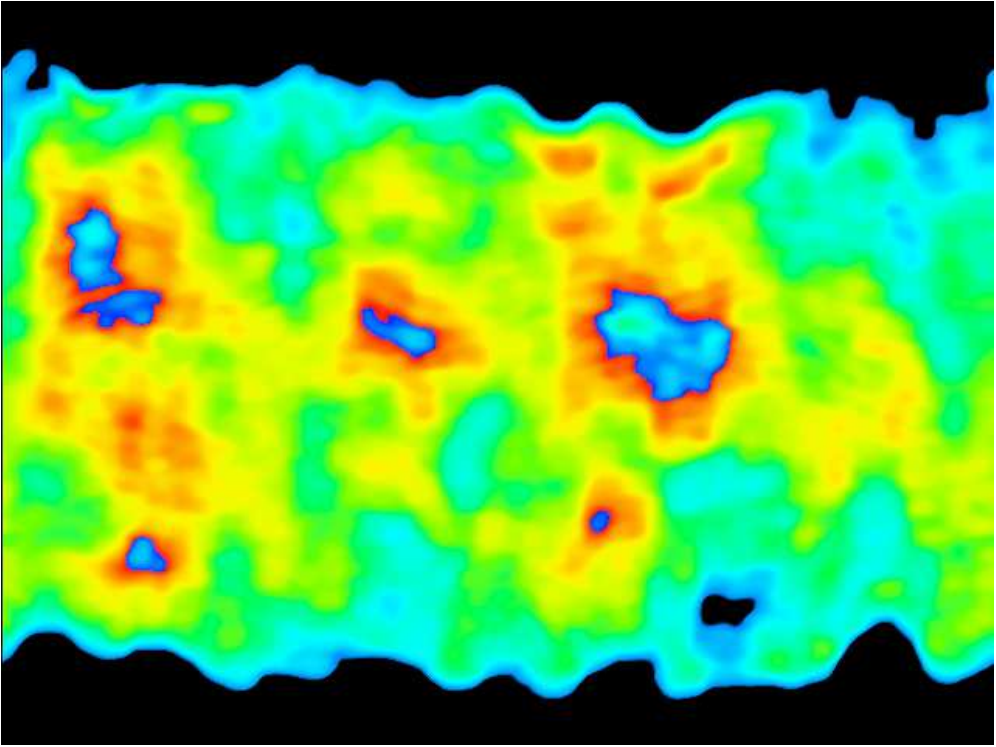}
\caption{One time step in a computer simulation of compressible magnetic
turbulence, with grey scale from black to white, and then black
to white again, representing increasing density. The structure is
hierarchical with at least 4 levels of clumps within
clumps visible here.}
\label{fig:sim} \end{figure}

We would like to know whether non-linear gas motions (i.e.
``turbulence'') with a magnetic field can produce structures and
velocities similar to what is observed in space. Figure \ref{fig:sim}
shows one time step in a computer simulation of a supersonically
turbulent, magnetic gas. The density is plotted
as a grayscale, with two cycles of gray representing the span of density
in the central ``cloudy'' region of a two-dimensional MHD simulation.
The initial conditions are a uniform density (value 1), and pressure
(value 1, giving a sound speed of 5/3), with a uniform magnetic field
oriented vertically in the figure. The initial Alfven speed for this
field is 10 in these units, which is supersonic. There is no
self-gravity. The full grid measures 800 cells in the vertical
direction, and 640 cells horizontally (only the central 480 cells in
the vertical direction are shown here). After this initial setup, the
magnetic field lines at the top and bottom of the grid are accelerated back and
forth, horizontally, with random accelerations for random durations at
random times, taking big swaths of field lines all at once. The
distribution of swath size is the same as we measured for the
distribution of interstellar cloud sizes, namely a power law with
$n(R)dR\propto R^{-3.3}dR$ (Elmegreen \& Falgarone 1996). What this result
simulates is the non-linear excitation of field lines by moving
neighboring clouds. 

The gas in the simulation has two stable thermal states, one at a
temperature corresponding to a sound speed of 1, and another at a
temperature corresponding to a sound speed of $10^{0.5}$ (i.e., 10 times
hotter). The magnetic field-line excitations at the top and bottom of
the grid heat up the gas there to the high temperature state, and also
drive the motion of gas toward the grid center. This forms a dense
``cloud'' at the vertical center of the grid, 
and this cloud takes the cool thermal
state. The simulation was run for $10^5$ timesteps, with each time step
equal to one-tenth of the time it takes an initial Alfven wave to travel
over a distance of one cell. 

The figure shows the dense central part of the grid, measuring 640 cells
horizontally and 480 cells vertically. Substructure inside the dense
part is apparently hierarchical, with some ``clumps'' containing
 4 levels of
substructure. The larger pieces move with a mildly supersonic speed, and
last for about one sound crossing time; during this time, they are
``bound'' by  ram pressure from their constant motion. That is, the
${\bf v}\bullet\nabla{\bf v}$ term in the equation of motion acts like a
binding pressure. The thermal pressure is actually lower between the
clumps than inside the clumps, so they are not bound by thermal
pressure; i.e., there is no warm interclump medium. 

Simulations like these suggest that hierarchical structure is a natural
consequence of non-linear magnetic wave interactions. This gives an indication
that perhaps turbulence is causing some of the structures seen
in interstellar clouds. We have a long way to go before the picture
is complete. 

\subsection{Density wave structure in clouds}

Optical images of galaxies like ours show such dominance by spiral arms
that we should always question whether any local interstellar
structures might have been caused by density waves.  Obviously, the
string of giant molecular clouds along the Sagittarius-Carina arm,
including the M17 and M16 clouds in the first quadrant, the eta Carina
cloud in the third quadrant, and many others (e.g., Cohen et al. 1985),
should look like pieces in a giant spiral arm {\it dust lane} if viewed from
outside the Galaxy.  There are probably streaming motions along this
dustlane too.

Spiral arm streaming motions (Roberts 1969) are well documented for
other galaxies (e.g., Kuno \& Nakai 1997) and for some regions in our
Galaxy, particularly along the tangent points of the inner spiral arms
(Burton 1992).  These streaming motions affect our estimates for the
kinematic distances to molecular clouds.   For example, the longitude
and velocity of the M17 cloud, $l=15^\circ$ and $v=20$ km s$^{-1}$,
give it a kinematic distance of 2.5 kpc in the standard Galactic
rotation model. But Hanson, Howarth \& Conti (1997) found a
spectroscopic distance in the range of 1.1 to 1.7 kpc.  At the
preferred distance of 1.3 kpc, its radial velocity should be only 10 km
s$^{-1}$. This implies that the M17 molecular cloud has a streaming
motion in the radial direction of $\sim10$ km s$^{-1}$, moving away
from the Sun. This is consistent with the expected sense of spiral wave
streaming, i.e., directed radially inward along the arm inside
corotation, or it corresponds to a purely azimuthal streaming of $\sim30$
km s$^{-1}$. Thus, the true streaming speed is somewhere between 10 and 30
km s$^{-1}$.  In any case, the observed radial speed gives it a
kinematic distance that is wrong by nearly a factor of 2!  Such an
error affects the mass estimate for M17 and
other clouds as well as our perception of Galactic
spiral structure.

\section{Further Implications of Hierarchical/Fractal Structure}

\subsection{Intercloud Medium}

If turbulence makes some clouds and cloud clumps, then it also makes
some of the low density regions between these clouds and clumps. These
are the spaces cleared out by the turbulent motions when the gas is
moved into denser regions. For a fractal cloud population, the
intercloud medium has certain regular properties that also come from
fractal geometry. 

One of these properties is the size distribution of holes. In
a hierarchical model with clumps inside other clumps, each level in the
hierarchy has a constant proportion of the number of empty regions of a
particular size to the number of clumps of that size. This means that
the hole size distribution is the same as the clump size distribution,
which is $n(S)d\log S \propto S^{-D}d\log S$ for size $S$ and fractal
dimension $D$ (Mandelbrot 1982). The dimension for this type of fractal
is defined by the relation $D=\log N/\log L$ for number of
subclumps in a clump, $N$, and for size ratio of clump to subclump $L$. It
may also be defined from the scaling of mass with size $M\propto S^D$.
If $D\sim2.3$ from the cloud size distribution and mass-size relation
(Elmegreen \& Falgarone 1996), then the hole size distribution should be
$S^{-2.3}d\log S$ too. 

The fractal Brownian motion model proposed by Stutzki et al. (1998)
also has a size distribution for clouds equal to the size
distribution for holes. This model begins with a white noise spectrum,
and then convolves this with a power law that has decreasing power at
high frequencies. The result is a noise model with the large scale
features emphasized. These features, like the original noise, are
equally likely to have positive or negative excursions around the mean
density. If we identify the positive excursions with clouds and cloud
clumps and the negative excursions with holes between the clouds, then
the size distribution for holes will be the same as the size
distribution for clumps. 

These predicted size distributions are steeper than the shell size
distribution found by Oey \& Clarke (1997), who got a power law with a
slope of $-1.3$ (instead of $-2.3$) on a $\log-\log$ plot. Oey \&
Clarke's objects are real shells, however, not just random regions with
locally low densities. There are no studies yet of the distribution of
hole sizes, but consideration of this might prove fruitful.

The volume filling factor of low density gas is another important
consideration for comparisons with observations. For the hierarchical
fractal model discussed above, the low density volume filling factor
equals  \begin{equation} f_{holes}=1-{\cal
C}^{\left(D/3\right)-1}\approx80\% \label{eq:ficm} \end{equation} for
density contrast ${\cal C}$ between the highest and lowest levels 
(Elmegreen 1997). To
evaluate this, we used ${\cal C}=10^3$ considering densities from 0.1 cm
$^{-3}$ to 100 cm$^{-3}$ in the diffuse medium, or from 100 cm$^{-3}$ to
$10^5$ cm$^{-3}$ in molecular clouds. The large value of this filling
factor indicates that {\it a turbulent interstellar medium should be mostly
empty, with most of the mass in the form of clouds that occupy only a
small fraction of the volume}. 

This result challenges the usual interpretation of the open frothy
structure of the interstellar medium as an indication of stellar winds
and supernovae (Brand \& Zealey 1975; Hunter \& Gallagher 1990). More
likely, the frothy structure is from a combination of these direct
stellar processes plus pervasive turbulent motions. For the turbulent
motions, the energy can be put into the gas in non-explosive forms
(e.g., shear, magnetic wave interactions, etc.), although some of it may
ultimately come from explosions around stars too. A recent fractal
interpretation of the frothy structure in HI maps of the SMC was made by
Stanimirovic et al. (1998).

Equation \ref{eq:ficm} gives the total filling factor of low density gas
from all levels in the hierarchy. There is another filling factor for a
hierarchical cloud distribution and that is the filling factor of
regions with a certain average gas density, $<\rho>$,
\begin{equation}f\left(<\rho>\right) d\log <\rho>\equiv S^3
n\left(R\left[<\rho>\right]\right) d\log S \propto
{{1}\over{<\rho>}}d\log <\rho>. \label{eq:fave} \end{equation} This
result is easily derived using the intermediate step shown, along with
the size distribution function $n(R)d\log R\propto R^{-D}d\log R$ and
the density-size relation $<\rho>\propto S^{D-3}$, which comes from the
mass-size relation that defines the fractal dimension $D$ in this model,
$M\propto S^D$. Equation \ref{eq:fave} indicates that the summed volume
of regions with a certain average density occupies a smaller and smaller
fraction of the total volume as that density increases. Equation
\ref{eq:ficm} above suggests further that as this average density
increases (and ${\cal C}\rightarrow1$), the gas inside each dense piece
becomes more and more uniform. Thus we recover the standard model of
{\it interstellar clouds composed of numerous, tiny, near-uniform
clumps, arranged in a fractal pattern with the average gas density
varying inversely with size} (for $D$ near 2).

We also get an explanation for the common perception that clumps in
emission line surveys generally occupy 1--10\% of the volume (e.g., Blitz
1993). Evidently, this is a selection effect caused by the
limitations of molecular emission maps which are typically sensitive
to only a factor of $\sim10-100$ range in density for any one survey.

\begin{figure} 
\vspace {6.in}
\includegraphics{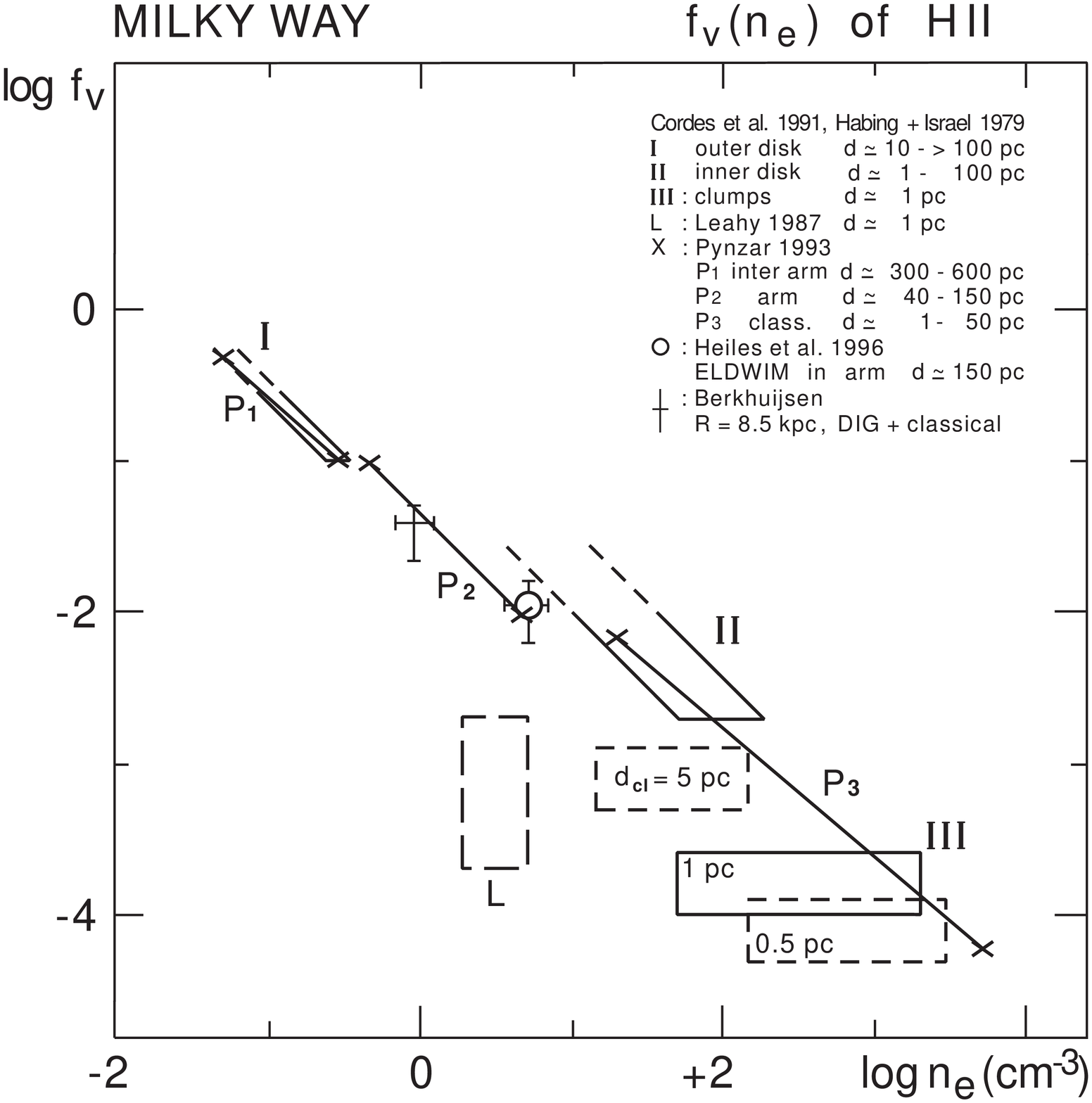}
\caption{ Volume filling factor of ionized gas in the Milky Way as a function of
density, showing the inverse-density relation typical for hierarchical
fractal structure. Figure from Berkhuijsen (1998).}
\label{fig:berk} \end{figure}

Berkhuijsen (1998) has obtained a remarkable result recently in plotting
the volume filling factor of ionized interstellar gas as a function of
its local average density. The result, shown in figure \ref{fig:berk},
is a $1/<\rho>$ distribution, as expected for a turbulent fluid with any
fractal dimension. This does not mean that the ionized fluid itself is
turbulent in this way; in most regions, the ionization structure is
probably just a reflection of the neutral structure of the gas before it
was ionized.

\subsection{Cloud and Intercloud Mean Free Paths}

Hierarchical or fractal structure in the intestellar medium implies that clouds
are highly clumped, i.e., that any one cloud is likely to have another cloud nearby, 
and, conversely, any empty region is not particularly likely to have a cloud nearby.
Such clumping makes the mean free path for intersecting a cloud on the line
of sight different for regions near clouds than for regions between clouds. 
Thus the {\it intercloud} mean free path can be larger than the {\it average} mean
free, which has always been obtained in the past by assuming a uniform density.  This 
result has 
important implications for the distance ionizing photons can travel once they
leave a molecular cloud/OB association complex. 

The mean free path is generally defined as
$\lambda=\left(nS^2\right)^{-1}$ for number density of clouds $n$ and
cross sectional area $S^2$. When $n$ is not uniform, but clumped
instead, this expression has a different value on lines of sight
near cloud complexes than it does on lines of sight between cloud
complexes. The hierarchical cloud model discussed above (Elmegreen 1997)
can be used to
show that the number of sub-clouds on a line of sight through a fractal
cloud is \begin{equation}N_{sub-clouds}={\cal
C}^{\left(D-2\right)/3}\sim2.5. \end{equation} If there are 8
interstellar absorption lines kpc$^{-1}$ on {\it average} (Blaauw 1952),
then these correspond to $8{\cal C}^{-\left(D-2\right)/3}\sim3$
fractal cloud complexes per kiloparsec. Indeed, most
of the local gas is not randomly distributed with a constant space
density of clouds everywhere. It is highly clumped 
into cloud complexes associated
with the OB associations in Orion, Sco-Cen, and Perseus, and with 
the Taurus clouds.
These are the types of clouds that occur with the rate of  
3 kpc$^{-1}$. 

The average mean free path on lines of sight {\it between} the fractal
cloud complexes should be ${\cal C}^{\left(D-2\right)/3}/8\sim
\left(3/{\rm kpc}\right)^{-1}\sim330$ pc. This may be defined as the
pure intercloud mean free path. It exceeds the Galactic scale
height, which means that once photons get into the disk intercloud
medium, they are likely to reach the halo. For this reason, the ionization
of the halo, giving the ``Reynolds'' layer (Reynolds 1995), may be much
easier than previously thought (compare this result to the uniform cloud
model by Miller \& Cox 1993).

\subsection{The Diffuse Ionized Gas}

HII regions expand away from giant molecular clouds into the intercloud
medium. For fractal clouds, the neutral density varies inversely with
distance $R$ as $\rho\propto R^{D-3}\sim R^{-0.7}$ for $D\sim2.3$. When
the ionization front reaches this neutral density, it heats and ionizes
the gas, causing the smallest pieces to expand and mix with each other.
This makes the local density more uniform on a timescale equal to the sound
crossing time for the small pieces, but it does not affect the overall
density gradient much because the sound crossing time is large on a
large scale. Thus a young HII region should have a radial density
gradient reflecting the $R^{D-3}$ gradient in the initial neutral cloud. As a
result, there should be an extended component of {\it diffuse ionized gas}
surrounding most HII regions. The fraction of the ionization outside the
conventional Str\"omgren sphere can be large, perhaps 50\% (Elmegreen
1997).  Such extended ionization around HII regions
presumably corresponds to the observations of diffuse ionized gas in
our Galaxy (Reynolds 1995) and other galaxies
(Walterbos \& Braun 1994; Ferguson et al. 1996). This diffuse ionized
gas is clearly associated with the outlying parts of individual HII regions in many cases (Walterbos \& Braun 1994).

\subsection{Cloud Mass Functions}

The mass distribution of purely hierarchical structures should 
scale as $Mn(M)d\log M=$ constant, or (Fleck 1996):
\begin{equation}n(M)dM\propto M^{-2}dM.
\label{eq:fleck}
\end{equation}
Actually, $n(M)\sim M^{-1.7-1.8}$ from clump surveys (Stutzki et al. 1998;
Heithausen et al. 1998). 

Elmegreen \& Falgarone (1997) got $n\sim M^{-\alpha_M}$ with
$\alpha_M\sim1.7-1.8$ from a model of hierarchical
cloud structure using the observed $n(S)dS\propto S^{-\alpha_S}dS$ for size
$S$ with $\alpha_S=1+D\sim3.3$, and the observed $M\propto S^\gamma$ with $\gamma\sim3$,
and then combining these relations with the standard conversion
$n(M)dM=n(S)dS$, which gives $\alpha_M=1+D/\gamma$.

Stutzki et al. (1998) got  $n(M)\sim M^{-1.8}$ with a cloud
model based on fractal Brownian motion, using $\alpha_M=3-\left(8-
2D\right)/\gamma$ and the same $D$ and $\gamma$.  The physical
difference between these two models is that hierarchical clouds
have all their small clumps inside larger clumps, whereas
fractal Brownian motion clouds have small structure with equal
probability everywhere, although it is weaker outside the larger clumps. 
In theory, the Brownian motion model makes more sense for turbulence, 
which should have some density structure everywhere,
but in practice, the low density regions outside of the main hierarchy in an interstellar
cloud may be ionized or invisible in CO, in which case only the dense hierarchical
core, which is common to both models, will be seen. 

The $M^{-2}$ function also applies to the probability of choosing part
of a hierarchical cloud from any level. If stars form at
all levels in a hierarchy because all levels have dense gas in 
tiny clumps, then the resulting stars will be hierarchically clumped
too.   Such a distribution for stars is in fact commonly observed
(Elmegreen \& Efremov 1997, 1998).  If, in addition, bound clusters
form with some high and nearly constant efficiency, 
so $M_{cluster}\propto M_{cloud\;piece}$, then the mass distribution
for star {\it clusters} will be about the same as the mass distribution
for cloud structure (Elmegreen \& Falgarone 1996).

Battinelli et al. (1994) found cluster mass functions from two samples
that had
power law slopes equal to $\alpha_{M,clusters}=-2.13\pm0.15$ and 
$-2.04\pm0.11$.  Elmegreen \& Efremov (1997) similarly found
$\alpha_{M,clusters}\sim2$ for three different age groups in the LMC. These
cluster mass functions are very similar to the expectations from 
hierarchical clouds, given by equation \ref{eq:fleck}.

\end{document}